\documentclass[twocolumn,aps,prb,amsmath,amssymb]{revtex4} 
\usepackage{bm}
\usepackage{enumitem}
\usepackage{graphicx}
\usepackage{color,xcolor}
\usepackage{amsmath}
\usepackage{amssymb}

\newcommand{\ud}{{\mathrm{d}}}

\newcommand{\B}{\mbox{\tiny B}}
\newcommand{\tS}{\mbox{\tiny S}}

\newcommand{\T}{\mbox{\tiny T}}

\newcommand{\greater}{\mbox{\tiny$>$}}
\newcommand{\lesser}{\mbox{\tiny$<$}}

\newcommand{\w}{\omega}
\newcommand{\ti}{\tilde}
\newcommand{\nl}{\nonumber \\}

\newcommand{\be}{\begin{equation}}
\newcommand{\ee}{\end{equation}}
\newcommand{\bsube}{\begin{subequations}}
\newcommand{\esube}{\end{subequations}}
\newcommand{\Eq}[1]{Eq.\,(\ref{#1})}
\newcommand{\Eqs}[1]{Eqs.\,(\ref{#1})}
\newcommand{\Fig}[1]{Fig.\,\ref{#1}}

\newcommand{\la}{\langle}
\newcommand{\ra}{\rangle}

\begin{document}

\title{Optimal control on open quantum systems and application to non-Condon photo-induced electron transfer}

\author{Zi-Fan Zhu}
\affiliation{Hefei National Laboratory,
University of Science and Technology of China, Hefei, Anhui 230088, China}
\affiliation{Hefei National Research Center for Physical Sciences at the Microscale and Department of Chemical Physics, University of Science and Technology of China, Hefei, Anhui 230026, China}

\author{Yu Su}
\affiliation{Hefei National Laboratory,
University of Science and Technology of China, Hefei, Anhui 230088, China}
\affiliation{Hefei National Research Center for Physical Sciences at the Microscale and Department of Chemical Physics, University of Science and Technology of China, Hefei, Anhui 230026, China}

\author{Yao Wang}
\email{wy2010@ustc.edu.cn}
\affiliation{Hefei National Laboratory,
University of Science and Technology of China, Hefei, Anhui 230088, China}
\affiliation{Hefei National Research Center for Physical Sciences at the Microscale and Department of Chemical Physics, University of Science and Technology of China, Hefei, Anhui 230026, China}

\author{Rui-Xue Xu}
\email{rxxu@ustc.edu.cn}
\affiliation{Hefei National Laboratory,
University of Science and Technology of China, Hefei, Anhui 230088, China}
\affiliation{Hefei National Research Center for Physical Sciences at the Microscale and Department of Chemical Physics, University of Science and Technology of China, Hefei, Anhui 230026, China}


\date{\today}

\begin{abstract}
In this work, we develop an optimal control theory on open quantum system and its environment, and exemplify the method  with the application to the non-Condon photo-induced electron transfer (PET) in condensed phase.
This method utilizes the dissipaton theory, proposed by Yan in 2014 for open quantum systems,
which provides an exact description of the dissipative system while also enables rigorous characterization and control of environmental hybridization modes, fully taking into account the non-perturbative and non-Markovian effects. 
Leveraging the advantage of the dissipaton phase-space algebra, we present in this communication the theoretical strategy for optimal control on both system and environment simultaneously.
The control protocol is successfully demonstrated on PET for the environment--targeted--control facilitated transfer.
This work sheds the light on manipulating open systems dynamics via polarized environment.
\end{abstract}

\maketitle

Control of molecular dynamics with light has been a central focus of theoretical and experimental research over the past four decades.\cite{Bru92257,Gor97601,Tan855013,
Bru86541,  Pei884950, Kos89201, Jud921500,Yan932320,Yan941094,Koh953360,Koh95133,Yan973471,Yan98191,Xu9910611,Xu046600,Bri10075008,Mag21010101}
Among various control schemes, optimal control has turned out to be a powerful and robust method, using tailored field to drive dynamic process to a desired target,
and has been successfully applied from gas phase to condensed phase.\cite{Yan932320,Koh953360,Jud921500,Xu046600,Bri10075008,Mag21010101}
Usually, optimal control is a standard problem of functional optimization under constraints.
The main challenge in condensed phase control is the theoretical description of  open quantum systems, where not only the system but also its environment (thermal bath) may under the control of external fields.\cite{Zha09820,Liu11931}

In 2014, Yan proposed an exact dissipaton formalism for open quantum systems,\cite{Yan14054105} which introduced a quasi-particle concept, the {\it dissipatons}, to establish a novel theoretical framework for characterizing and manipulating environmental collective dynamics and statistical properties.
With the aid of the quasi-particle algebra,\cite{Wan20041102} not only is the dissipaton theory
convenient for bath collective dynamics and polarizations under fields,\cite{Zha15024112,Che21244105} but also is it straightforward for extension to nonlinear bath couplings which is of non-Gaussian statistics.\cite{Xu18114103,Zhu25234103,Su23024113,Su254107} 
Besides, the dissipaton-equation-of-motion (DEOM) can be constructed to compute both real-time dynamics and imaginary-time evolution as well as non-equilibrium
thermodynamic properties.\cite{Gon20154111, Wan22044102, Wan22170901}

This work aims at the optimal control on the dynamics of open quantum systems via the DEOM simulations which offers exact treatments 
in a systematic way 
in the condition that not only the system but also the environment interact with the light.
After elaborating the theoretical scenario,  we will carry out demonstrations on the control of
non-Condon
photo-induced electron transfer (PET) 
dynamics.
PET is a fundamental process governing charge separation in natural and artificial systems, from photosynthetic reaction centers to molecular electronics.\cite{Dad1610212}
It is found that the PET dynamics may be sensitive to environmental fluctuations.\cite{Dor132746,Cre13253601}

One key challenge in controlling such ultrafast processes 
is that both electronic and environmental degrees of freedom may  participate.
This kind of optimal control strategy will be developed in this work, exploiting the dissipaton phase-space 
description on environmental
polarized dynamics under the control field.
As an exact, non-Markovian quasi-particle encoder of environmental hybridization dynamics,
the 
dissipatons--incorporated optimal control 
strategy
provides  systematic and precise  control involving environmental polarized dynamics in non-equilibrium regimes.

Let us start from the total Hamiltonian
$H_{\T}(t)=H_M+H'(t)$ 
where the matter Hamiltonian is  decomposed into the system-plus-environment (bath) form as 
\be\label{HM}
H_{M}=H_{\tS}+H_{\tS\B}+h_{\B}\quad {\rm with} \quad H_{\tS\B}=\sum_mQ_mF_m.
\ee
Here $\{Q_m\}$ and $\{F_m\}$ are the hybridized system and bath operators, respectively.
The matter-field interaction reads
$H'(t)=-\hat \mu_{\T}\varepsilon(t)$ where $\varepsilon(t)$ is the classical external field.
 The total dipole operator
assumes the form in the Herzberg–Teller approximation as
\be\label{totdipole}
\hat \mu_{\T}=\hat \mu_{\tS}\big(1+\sum_mv_mX_m\big)\ee 
where $\{X_m\}$ are coordinates of bath modes.
Throughout the paper, we set the Planck constant and Boltzmann constant as units ($\hbar=1$ and $k_B=1$), and $\beta=1/T$, with $T$ being the temperature.

For Gaussian bath, in the microscopic level,
$h_{\B}$ is composed of harmonic oscillators and 
$\{F_m\}$ linearly depend on 
$\{X_m\}$.
 The influence of bath is entirely described by the bath correlation functions averaged in the canonical bath space, $\la F^{\B}_m(t)F^{\B}_{m'}(0)\ra_{\B}$ with $F^{\B}_{m}(t)\equiv e^{ih_{\B}t}F_{m}e^{-ih_{\B}t}$, and related to the bath spectral densities $J_{mm'}(\w)$ 
 via the fluctuation--dissipation theorem as\cite{Wei21,Yan05187} 
\be\label{fdt}
  \la F^{\B}_m(t)F^{\B}_{m'}(0)\ra_{\B}=\frac{1}{\pi}\int_{-\infty}^{\infty}\!{\rm d}\w\,
  \frac{e^{-i\w t}J_{mm'}(\w)}{1-e^{-\beta\w}}.
\ee
The bath contribution to dipole in \Eq{totdipole}
is assumed 
along the bath hybridization mode. 
It can be  function of other collective coordinate of bath, as long as its associated statistical property such as spectral density or correlation function is known.
The bath mode reorganization energy can be obtained via
\be
  \lambda_{m}=\frac{1}{2\pi}\int_{-\infty}^{\infty}\!{\rm d}\omega\, \frac{J_{mm}(\omega)}{\omega}.
\ee
By introducing
\be\label{omgb2}
\lambda_{m}\Omega^2_{m}
=\frac{1}{2\pi}\!\int_{-\infty}^{\infty}\!{\rm d}\w\,\w J_{mm}(\w)
\ee
to determine the characteristic frequency of the hybridized bath mode,
we have the relation for the phase-space 
coordinate
of the bath mode
\be\label{XmFm}
X_{m}=(2\lambda_m\Omega_{m})^{-\frac{1}{2}} F_{m}
\ee
and the conjugated momenta via
$\Omega_{m}P_{m}
=i[H_{M},X_{m}]$.

Adopt exponential series expansion for bath correlations 
\be\label{Ctexpan}
  \la F^{\B}_m(t)F^{\B}_{m'}(0)\ra_{\B}\simeq\sum^{K}_{k=1}\eta_{mm'k}e^{-\gamma_{k}t}.
\ee
This can be obtained via sum-over-pole decomposition of the Fourier integrand in \Eq{fdt} and contour integral,\cite{Hu10101106,Din12224103} certain minimum ansatz,\cite{Din16204110} or time--domain fitting scheme.\cite{Che22221102} 
The time-reversal relation is 
\begin{align} \label{FBt_corr}
\la F^{\B}_{m'}(0)F^{\B}_{m}(t)\ra_{\B}
=
\la F^{\B}_{m}(t)
F^{\B}_{m'}(0) \ra^{\ast}_{\B}
=
\sum^{K}_{k=1}\eta_{mm'\bar k}^{\ast} e^{-\gamma_{k} t}.
\end{align}
The exponents $\{\gamma_{k}\}$ in \Eqs{Ctexpan} and (\ref{FBt_corr})  
must be either real or complex conjugate paired, and $\bar k$ is determined in the index set $ \{k=1,2,...,K\}$
by $\gamma_{\bar k}=\gamma_{k}^{\ast}$.
Here, $\gamma_k$ runs over all involved exponents  but with $\eta_{mm'k}$ or
$\eta^{\ast}_{mm'\bar k}$  being zero if not really among the terms.

The dissipaton theory is established 
on basis of the \emph{dissipatons decomposition} that reproduces the correlation functions in \Eqs{Ctexpan} and (\ref{FBt_corr})
by introducing a number of dissipaton operators, $\{\hat{f}_{m k} \}$, such that \begin{align}\label{hatFB_in_f}
F_{m}=\sum^K_{k=1}  \hat{f}_{m k},
\end{align}
with ($\hat{f}^{\B}_{mk}(t)\equiv e^{ih_{\B}t}\hat{f}_{mk}e^{-ih_{\B}t}$)
\bsube\label{eq9}
\begin{align}\label{fx_corr}
\la \hat{f}^{\B}_{m k}(t)\hat{f}^{\B}_{m'j}(0)\ra_{\B}=\delta_{k j}\eta_{mm'k} e^{-\gamma_{k}t},
\\ \label{fx_corr_nu_rev}
\la \hat{f}^{\B}_{m'j}(0)\hat{f}^{\B}_{m k}(t)\ra_{\B}=\delta_{k j} \eta_{mm'\bar k}^{\ast} e^{-\gamma_{k}t}.
\end{align}
\esube
Now define the dissipaton density operators
(DDOs), which serve as
the dynamical variables in DEOM, as
\be \label{DDO}
  \rho^{(n)}_{\bf n}(t)
\equiv {\rm tr}_{\B}\Big[
  \big(\prod_{m k}\hat{f}_{m k}^{n_{m k}}\big)^{\circ}\rho_{\T}(t)
 \Big].
\ee
Here, $n=\sum_{m k}n_{m k}$, with $n_{m k}\geq 0$
for the bosonic dissipatons.
The product of dissipaton operators inside $(\cdots)^\circ$
is \emph{irreducible}, satisfying
$(\hat{f}_{m k}\hat{f}_{n j})^{\circ}
=(\hat{f}_{n j}\hat{f}_{m k})^{\circ}$
for bosonic dissipatons.
Each $n$--particles DDO, $\rho^{(n)}_{\bf n}(t)$, is labelled with
an ordered set of indexes, ${\bf n}\equiv \{n_{m k}\}$.
Denote for later use  ${\bf n}^{\pm}_{m k}$ differing from ${\bf n}$ only
at the specified dissipatons with their occupation numbers
$\pm 1$.
The zeroth-tier DDO, 
$\rho_{\bf 0}^{(0)}(t)=\rho_{0\cdots 0}^{(0)}(t)={\rm tr}_{\B}
\rho_{\T}
=\rho_{\tS}(t)$ is just the reduced system density operator.
The equation of motion for DDOs, i.e.\ the DEOM formalism, under the interaction of external field [cf.\,\Eqs{totdipole} and (\ref{XmFm})] can be finally obtained as\cite{Che21244105}
\begin{align} \label{gen_DEOM}
\dot\rho^{(n)}_{\bf n} \!
= & - \Big[i{\cal L}(t) + \sum_{m k}n_{m k}\gamma_{k}
\Big]
\rho^{(n)}_{\bf n} 
-i\sum_{m k} 
{\cal A}_m(t)
\rho^{(n+1)}_{{\bf n}^{+}_{m k}}
\nl &
-i\sum_{m k} 
n_{mk}
{\cal C}_{mk}(t)
\rho^{(n-1)}_{{\bf n}^{-}_{mk}}
.
\end{align}
Here, ${\cal L}(t) O \equiv 
[H_{\tS}-\hat\mu_{\tS}\varepsilon(t),  O]$, ${\cal A}_m(t)  O \equiv [ \ti Q_m(t),  O]$, and ${\cal C}_{m k}(t) O \equiv \sum_{m'}[\eta_{mm'k}  \ti Q_m(t) O
- \eta^{\ast}_{mm'\bar k} O \ti Q_m(t)]$, with $\ti Q_m(t) \equiv Q_m-v_m(2\lambda_m\Omega_m)^{-\frac{1}{2}}\hat \mu_{\tS}\varepsilon(t)$.
For later use in the control scenario, we elaborate more about the dissipaton phase-space algebra.\cite{Wan20041102} For the coordinate, $X_{m}=(2\lambda_m\Omega_{m})^{-\frac{1}{2}}  \sum_k \hat{f}_{m k}$, we have
\be
\label{genwicks}
\begin{split}
\rho^{(n)}_{\bf n}(t;\hat f_{mk}^{\greater})
&=\rho_{{\bf n}^{+}_{mk}}^{(n+1)}
 +\sum_{m'} n_{m'k} \eta_{m'mk}\rho_{{\bf n}^{-}_{m'k}}^{(n-1)},
\\
\rho^{(n)}_{\bf n}(t;\hat f_{mk}^{\lesser})
&= \rho_{{\bf n}^{+}_{mk}}^{(n+1)}
 +\sum_{m'}n_{m'k} \eta^{\ast}_{m'm \bar k}\rho_{{\bf n}^{-}_{ m'k}}^{(n-1)};
\end{split}
\ee
while for the momentum, $P_{m}=(2\lambda_m\Omega^3_{m})^{-\frac{1}{2}}  \sum_k {\gamma_{k}}\hat\varphi_{mk}$,
\be
\label{genwicks_momentum}
\begin{split}
\rho^{(n)}_{\bf n}(t;\hat \varphi_{mk}^{\greater})
&=-
 \rho^{(n+1)}_{{\bf n}^{+}_{mk}}
 +\sum_{m'}n_{m'k}\eta_{m'mk}\rho^{(n-1)}_{{\bf n}^{-}_{m'k}},
\\
\rho^{(n)}_{\bf n}(t;\hat \varphi_{mk}^{\lesser})
&=-
 \rho^{(n+1)}_{{\bf n}^{+}_{mk}}
 +\sum_{m'}n_{m'k}\eta_{m'm\bar k}^{\ast}\rho^{(n-1)}_{{\bf n}^{-}_{m'k}}.
\end{split}
\ee
We thus finish the outline of the DEOM theory for open quantum systems.

Turn now to the optimal control on quantum dynamics in condensed phase. The control objective is to find a form of external field, $\varepsilon(t)$, to optimize an expectation, $A(t_f)={\rm Tr}[\hat A\rho_{\T}(t_f)]$, at a time $t_f$,
where $\hat A$
is the target operator.
The optimal control theory 
in condensed phase has been systematically developed by Yan and co-workers 
in 
Ref.\,\onlinecite{Yan932320}. 
Generally the control field is resolved 
in an iterative way by a 
self-consistent functional equation.
In case the target state does not overlap with the initial state, people can compromise with the weak field scenario 
which is resolved via\cite{Yan932320}
\be \label{Meign}
\int_{t_0}^{t_f}\!{\rm d}\tau'\,{\cal M}(\tau,\tau')\varepsilon(\tau')=\Lambda\varepsilon(\tau),
\ee
with
\be\label{Mtau}
{\cal M}(\tau,\tau')=-{\rm Tr}[\hat A{\cal G}_M(t_f-\tau){\cal D}{\cal G}_M(\tau-\tau'){\cal D}\rho_{\T}(t_0)].
\ee
Here ${\cal G}_M(t)\hat O=e^{-iH_Mt}\hat Oe^{iH_Mt}$
and ${\cal D}\hat O=[\hat \mu_{\T},\hat O]$. Equation (\ref{Mtau}) satisfies ${\cal M}(\tau,\tau')={\cal M}(\tau',\tau)$ in the condition $[H_M,\rho_{\T}(t_0)]=[\hat A,\rho_{\T}(t_0)]=0$.
Adopting an equally spaced time-grid representation, $\varepsilon(t)$ becomes a vector, while $\mathcal{M}(\tau,\tau')$ a symmetric matrix. 
Thus, \Eq{Meign} becomes an eigenvalue
equation. 
The optimal control field is then obtained as the eigenvector with the eigenvalue,
$\Lambda=A^{(2)}(t_f)/(\frac{I}{2})$,
where $I\equiv\int_{t_0}^{t_f}\!{\rm d}t\,|\varepsilon(t)|^2$ is the integration strength.\cite{Yan932320}

Evaluation on \Eq{Mtau} via the dissipaton formalism goes with the following steps:
\begin{enumerate}

\item Determine the steady-state correspondence
of $\rho_{\T}(t_0)\Rightarrow
\big\{\rho^{(n)}_{{\bf n}}(t_0)\big\}$. It can be  evaluated as the solution to $\dot\rho^{(n)}_{\bf n}=0$ of the field-free \Eq{gen_DEOM}.



\item The correspondence of ${\cal D}\rho_{\T}(t_0)\Rightarrow\{\rho^{(n)}_{\bf n} (t_0;{\cal D})\}$
is
\begin{align} \label{gen_dipo}
&\quad\rho^{(n)}_{\bf n}
(t_0;{\cal D})\!
= 
{\cal D}_{\tS}
\rho^{(n)}_{\bf n} 
+\sum_{m k} 
\ti v_m{\cal D}_{\tS}
\rho^{(n+1)}_{{\bf n}^{+}_{m k}}
\nl & \quad \quad 
+\sum_{m m'k} 
\ti v_{m} n_{m'k}\Big(\eta_{m'mk} \hat \mu_{\tS}\rho^{(n-1)}_{{\bf n}^{-}_{m'k}}
-
\eta^{\ast}_{m'm\bar k}\rho^{(n-1)}_{{\bf n}^{-}_{m'k}}\hat \mu_{\tS}\Big),
\end{align} 
denoting ${\cal D}_{\tS}\hat O\equiv [\hat \mu_{\tS},\hat O]$
and $\ti v_m\equiv v_m(2\lambda_m\Omega_m)^{-\frac{1}{2}}$.

\item Perform the field-free DEOM propagation from $\{\rho^{(n)}_{\bf n}
(t_0;{\cal D})\}$ for a time period $t=\tau-\tau'$ to obtain $\{\rho^{(n)}_{\bf n}
(t;{\cal D})\}$ which corresponds to ${\cal G}_M(\tau-\tau'){\cal D}\rho_{\T}(t_0)$. 

\item Obtain ${\cal D}{\cal G}_M(\tau-\tau'){\cal D}\rho_{\T}(t_0)\Rightarrow\{\rho^{(n)}_{\bf n}(t;{\cal D}^2)\}$ similarly as Step 2,
with the {\it l.h.s.}\ of \Eq{gen_dipo} is 
$\rho^{(n)}_{\bf n}(t;{\cal D}^2)$ and the DDOs enter the {\it r.h.s.}\ of \Eq{gen_dipo} are from $\{\rho^{(n)}_{\bf n}(t;{\cal D})\}$.

\item Perform again the field-free DEOM propagation from $\{\rho^{(n)}_{\bf n}(t;{\cal D}^2)\}$ for duration time $t' = t_f-\tau$, and obtain ${\cal G}_M(t_f-\tau){\cal D}{\cal G}_M(\tau-\tau'){\cal D}\rho_{\T}(t_0)\Rightarrow\{\rho^{(n)}_{\bf n}(t',t;{\cal D}^2)\}$. 

\item Finally, obtain ${\cal M}(\tau, \tau')$ in \Eq{Mtau}. This step may be case by case. Some details will be exemplified in numerical demonstration.

\end{enumerate}

\begin{figure}
\includegraphics[width=0.9\columnwidth]{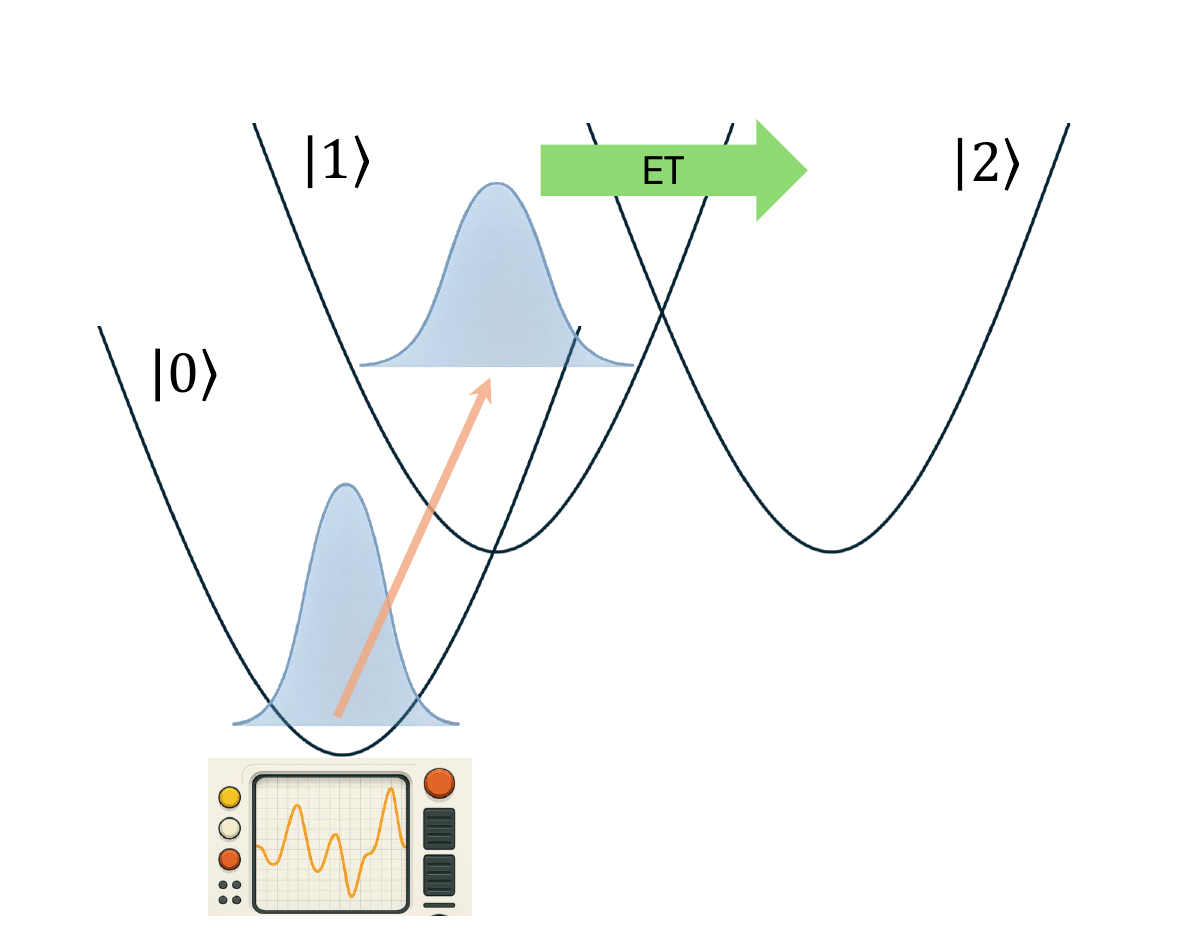}
\caption{Sketch of the PET system in this work.}\label{fig1}
\label{sketch}
\end{figure}

For numerical demonstration, we focus on the non-Condon
photo-induced electron transfer (PET) reaction
in a ground-donor-acceptor system that is embedded in a solvent. 
Before the external field action, the system is initially at the ground state ($|0\ra$), thermally equilibrated with the solvent. Upon the pulsed control field turned on, the system is prompted to the donor state ($|1\ra$). It then follows with the subsequent  transfer to the acceptor state ($|2\ra$). 
The whole process is illustrated in \Fig{sketch}.
The total matter Hamiltonian is
\be\label{HMet}
  H_{M}=\sum_{m=0}^2(\epsilon_{m}+h_{m})|m\ra\la m|+\hat V_{12}.
\ee
Here, $\epsilon_{m}$ is the electronic-state energy, and $h_m$ is the Hamiltonian of the solvent according to each state. The transfer coupling term is $\hat V_{12}=V(|1\ra\la 2|+|2\ra\la 1|)$ with $V$ being the interstate coupling strength. 
The system's contribution to  dipole in \Eq{totdipole} is
$\hat \mu_{\tS}=u(|0\ra\la 1|+|1\ra\la 0|)$.
As the system is initially located at the ground state $|0\ra$ in thermal equilibrium with the solvent $h_0$, we select $h_0$ as the referenced bath Hamiltonian $h_{\B}=h_{0}=\frac{1}{2}\sum_j \omega_j \left(p_j^2+x_j^2\right)$. The solvent equilibrium positions are displaced, according to the donor and acceptor states $|1\ra$ and $|2\ra$, i.e.,
$
  h_m=\frac{1}{2}\sum_j \omega_j \left[p_j^2+(x_j+d_{mj})^2\right];m=$1,2.
It leads to 
$\delta h_m=h_m-h_0=\lambda_m+F_{m}$ where $\lambda_m =\frac{1}{2}\sum_j \omega_j d_{mj}^2$ and $F_{m}=\sum_j \omega_jd_{mj}  x_j$,
with the spectral density,
$J_{mm'}(\omega>0)=\frac{\pi}{2}\sum_j\omega_j^2d_{mj}d_{m'j}
  \delta(\w-\w_j)=-J_{m'm}(-\omega)$.
These constitute the microscopic foundation of \Eqs{HM}--(\ref{XmFm}) 
for PET.
Separating the total matter Hamiltonian [\Eq{HMet}] into the system-plus-bath form [\Eq{HM}], 
we have ($h_{\B}=h_{0}$,
$\delta\epsilon_{m0}=\epsilon_m-\epsilon_0$)
\bsube
\begin{align}
 H_{\tS}
&=\sum_{m=1}^{2}(\delta\epsilon_{m0}+\lambda_m)|m\ra\la m|+\hat V_{12},\\
 H_{\tS\B}
&=\sum_{m=1}^{2}|m\ra\la m|F_m\equiv\sum_{m=1}^{2}Q_mF_m.
\end{align}
\esube
With the electronic-state transition, the solvent modes here involve only linear displacements.
In reality, there may be also frequency change and Duschinsky rotation. These complexities can be treated by the method in Ref.\,\onlinecite{Zhu25234103}.   

The electron transfer (ET) reaction is prompted by field excitation from the initially thermalized ground state, $\rho_{\T}(t_0)=\rho_{\B}^{\rm eq}|0\ra\la 0|$ with $\rho_{\B}^{\rm eq}=e^{-\beta h_{\B}}/{\rm tr}_{\B}e^{-\beta h_{\B}}$
which corresponds to $\rho_{\bf 0}^{(0)}(t_0)=1$ while $\rho^{(n>0)}_{\bf n}(t_0)=0$.
The control target operators are chosen as
\be\label{target}
\hat A=
\ti\rho_1(\ti\beta)|1\ra \la 1|
\ee
where
\be\label{target2}\ti\rho_1(\ti\beta)=\frac{e^{-\ti\beta 
\ti H_1}}{{\rm tr}_1(e^{-\ti\beta 
\ti H_1})}
=2\sinh(\ti\beta\Omega_1/2) e^{-\ti\beta 
\ti H_1}, 
\ee
with
$\ti\beta$ being a pre-selected inverse temperature for the targeted wavepacket of the solvation coordinate $X_1$ on the donor state $|1\ra$ and 
\begin{align}\label{tiHm}
\ti H_1 
\!=\!
\frac{\Omega_1}{2}\big[
 P^2_1+(X_1+D_1)^2 \big];\ \,
D_1
\!=\!
\Big(\frac{2\lambda_1}{\Omega_1}\Big)^{\frac{1}{2}}.
\end{align}
In evluating \Eq{Mtau}, the final step now goes by introducing $e^{-\ti\beta 
\ti H_1}{\cal G}_M(t_f-\tau){\cal D}{\cal G}_M(\tau-\tau'){\cal D}\rho_{\T}(t_0)\Rightarrow\vartheta^{(n)}_{\bf n}(\ti\beta)$, which satisfies the differential equation \begin{align}\label{theta}
\frac{\ud}{\ud{\ti\beta}}\vartheta^{(n)}_{\bf n}(\ti\beta) = -\vartheta^{(n)}_{\bf n}(\ti\beta;{\ti H_1}^{\greater}),
\end{align}
with the boundary condition $
\vartheta^{(n)}_{\bf n}(\ti\beta=0)\!=\!\rho^{(n)}_{\bf n}(t',t;{\cal D}^2)
$.
Solving \Eq{theta} by substituting 
\Eq{tiHm} with using the dissipatons algebra for the involved operations (see details in 
Supplementary
Material), 
we can obtain 
$\{
\vartheta^{(n)}_{\bf n}(\ti\beta)\}$
for various $\ti\beta$ values and finally 
evaluate ${\cal M}(\tau, \tau')$ as
\begin{align}
{\cal M}(\tau, \tau')=2\sinh(\ti\beta\Omega_1/2)
\la 1|
\vartheta^{(0)}_{\bf 0}(\ti\beta)  |1\ra.
\end{align}

In the demonstration, the ET system parameters are selected as $\delta \epsilon_{10}=\delta \epsilon_{20}=4V=1$, in unit of $T$.
The bath spectral densities, $\{J_{mm'}(\w)\}$, adopt the form of Brownian oscillator \cite{Din12224103}
\be
J_{mm'}(\w)=
 {\rm Im}
\frac{2\Lambda_{mm'}\Omega^2}{
\Omega^2 - \w^2-i\w\zeta(\w)},
\ee
with $\zeta(\w) =\eta\Gamma/(\Gamma - i\w)$ being the solvent friction resolution function. 
We choose
$\Omega=0.4$, $\eta=0.8$, $\Gamma=3$, 
and
\be 
[\Lambda_{mm'}]=\left[\begin{array}{cc}
\lambda_1 & \lambda_1+\delta  \\
\lambda_1+\delta & \lambda_1+\lambda_{U}+2\delta
\end{array} \right].
\ee
Here, $\lambda_{U}$ is the ET renormalization energy and  $U\equiv F_2-F_1$ denotes the ET reaction coordinate.
The parameter $\delta$ characterizes the cross correlation between the control mode $F_1$ and $U$, with 
$\delta=0$ and $(\lambda_1 \lambda_{U})^{\frac{1}{2}}$ for the uncorrelated and fully-correlated conditions, respectively.
We select $\lambda_1=0.2$ and $\lambda_{U}=1.8$, and
$v_1=0.5$ and $v_2=0$ in \Eq{totdipole} for the PET.

\begin{figure}
\includegraphics[width=\columnwidth]{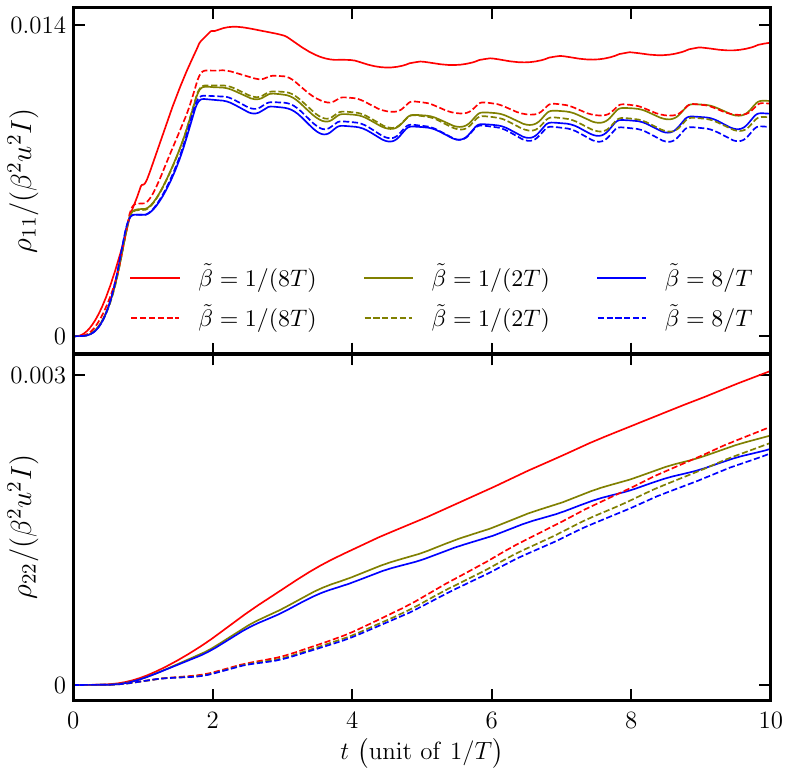}
\caption{Population evolutions of donor ($|1\ra$, upper panel) and acceptor ($|2\ra$, lower panel) states. The solid and dashed curves represent the  uncorrelated and fully-correlated cases, respectively. }\label{fig2}
\end{figure}

\begin{figure}
\includegraphics[width=0.95\columnwidth]{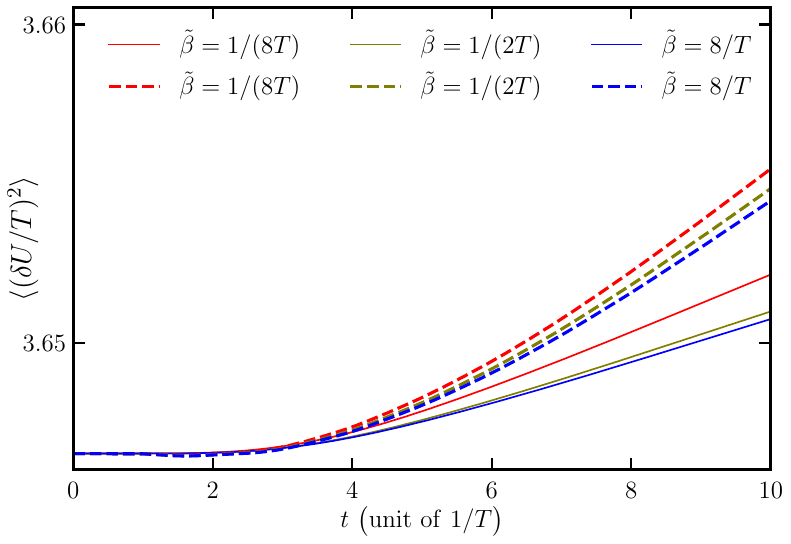}
\caption{The variance of  distribution of the reaction coordinate, with the solid and dashed curves representing the uncorrelated and  fully-correlated cases, respectively.}\label{fig3}
\end{figure}

Figure \ref{fig2} depicts the controlled PET evolution in the linear response regime, with varied values of $\ti \beta$ [cf.\,\Eqs{target} and (\ref{target2})].
The control target time is chosen as $t_f=1$ in unit of $T^{-1}$, and the obtained optimal control field is repeatedly applied. See details and shape of control fields in 
Supplementary
Material.
With the control targeted inverse temperature $\ti \beta\equiv 1/\ti T$ varying from $8/T$ to $1/(8T)$, the transfer evolutions as well as those depicted in other figures exhibit monotonic change in either uncorrelated or fully-correlated scenarios, due to our calculations across  a wide range of $\ti T$.
 In \Fig{fig2} we just pick the results for three target temperatures.
Apparently, higher target temperatures facilitate the ET process in the present setting.

From \Fig{fig2}, we also observe that ET is more facilitated in the uncorrelated condition compared to the fully-correlated scenario.
To indicate the behavior of the wavepacket of ET reaction coordinate $U$ during the  process under control, 
we plot the evolution of the variance $\la (\delta U)^2\ra$
in \Fig{fig3}. Similarly as in \Fig{fig2}, the targeted $\ti\beta$ plays relatively more important roles when control mode $F_1$ and  reaction coordinate $U$ are statistically uncorrelated, in comparison to the fully-correlated results. 
Notably, in correlated cases, the increased wavepacket width leads to the decrease in ET rate, compared to the uncorrelated cases at the same target temperature.
The controlled behavior of $F_1$ is shown in Supplementary
Material, together with the formulas to evaluate the variances 
$\la(\delta U)^2\ra$
and
$\la(\delta F_1)^2\ra$.

To summarize, in this communication we present a theoretical strategy for non-Condon optimal control of photo-induced reaction in condensed phase, leveraging the exact dissipaton-equation-of-motion (DEOM) formalism. By exploiting the phase-space dissipaton algebra, the developed optimal control protocol simultaneously governs not only electronic transitions but also solvation dynamics.
Numerical demonstrations successfully showcase the outcome of the targeted control and the role of correlation between hybridized system-bath modes.
This work sheds the light on manipulating open systems dynamics via polarized environment.
The strong-field optimal control involving iterative self-consistent algorithm  is currently in progress.

Support from  the National Natural Science Foundation of China (Grant Nos.\  22173088, 22373091, 224B2305), and the Innovation Program for Quantum Science and Technology (Grant No.\ 2021ZD0303301) is gratefully acknowledged. 
Simulations were performed on the robotic AI-Scientist platform of Chinese Academy of Sciences.

\end{document}


\title{
	Supplementary materials for ``Optimal control on open quantum systems and application to non-Condon photo-induced electron transfer''
}
\author{Zi-Fan Zhu}
\affiliation{Hefei National Laboratory,
University of Science and Technology of China, Hefei, Anhui 230088, China}
\affiliation{Hefei National Research Center for Physical Sciences at the Microscale and Department of Chemical Physics, University of Science and Technology of China, Hefei, Anhui 230026, China}

\author{Yu Su}
\affiliation{Hefei National Laboratory,
University of Science and Technology of China, Hefei, Anhui 230088, China}
\affiliation{Hefei National Research Center for Physical Sciences at the Microscale and Department of Chemical Physics, University of Science and Technology of China, Hefei, Anhui 230026, China}

\author{Yao Wang}
\email{wy2010@ustc.edu.cn}
\affiliation{Hefei National Laboratory,
University of Science and Technology of China, Hefei, Anhui 230088, China}
\affiliation{Hefei National Research Center for Physical Sciences at the Microscale and Department of Chemical Physics, University of Science and Technology of China, Hefei, Anhui 230026, China}

\author{Rui-Xue Xu}
\email{rxxu@ustc.edu.cn}
\affiliation{Hefei National Laboratory,
University of Science and Technology of China, Hefei, Anhui 230088, China}
\affiliation{Hefei National Research Center for Physical Sciences at the Microscale and Department of Chemical Physics, University of Science and Technology of China, Hefei, Anhui 230026, China}
\date{\today}

\begin{abstract}
\vspace{2em}
    These supplementary materials contain (I) Details on Eq.\,(23) of main text; and (II) More numerical details including shape of control field, behavior of variance of $F_1$, and some related formulas.
    \vspace{2em}
\end{abstract}

\maketitle
\onecolumngrid

\section{Details on Eq.\,(23) of main text}

Here, we compute $\vartheta^{(n)}_{\bf n}(\ti\beta;{\ti H_1}^{\greater})$ in Eq.\,(23) of main text. Since
\begin{align}
\ti H_1 =& \frac{\Omega_1}{2}\Big[
P_1^2 + ( X_1 + D_1)^2
\Big]=
\frac{\Omega_1}{2}\Big(
P_1^2 +  X^2_1\Big)
+ \frac{1}{2}\Omega_1 D^2_1 
 + \Omega_1 D_1 X_1 
=
\frac{\Omega_1}{2}\Big(
P_1^2 +  X^2_1\Big) 
+ \lambda_1
 + F_1, 
\end{align}
we need to compute the contributions of $\vartheta^{(n)}_{\bf n}\Big(\ti\beta;F_1^{\greater}\Big)$ and $\vartheta^{(n)}_{\bf n}\Big(\ti\beta;\frac{\Omega_1}{2}(
P_1^2 +  X^2_1)^{\greater}\Big)$, apart from the contribution of the constant $\lambda_1$. 
The $\vartheta^{(n)}_{\bf n}\Big(\ti\beta;F_1^{\greater}\Big)$ can be obtained directly via Eqs.\,(9) and (13) in the main text, while $\vartheta^{(n)}_{\bf n}\Big(\ti\beta;\frac{\Omega_1}{2}(
P_1^2 +  X^2_1)^{\greater}\Big)$ is in details as follows.

By using Eqs.\,(13) and (14) in the main text for the first-time action, we obtain 
\begin{align}
&\ \vartheta^{(n)}_{\bf n}\Big(\ti\beta;\frac{\Omega_1}{2}(
P_1^2 +  X^2_1)^{\greater}\Big)
=
\frac{\Omega_1}{2}
\Big[
\vartheta^{(n)}_{\bf n}(\ti\beta;(X^2_1)^{\greater}  )
+
\vartheta^{(n)}_{\bf n}(\ti\beta; (P^2_1)^{\greater})\Big]
\nl
& =
\frac{1}{2}
{\sqrt\frac{\Omega_1}{2\lambda_1}}
\!\sum_{k}
\Big[
\vartheta_{{\bf n}^{+}_{1k}}^{(n+1)}
(\ti\beta;X_1^{\greater})
 +\!\!\sum_{m=1}^2 n_{mk} \eta_{m1k}\vartheta_{{\bf n}^{-}_{mk}}^{(n-1)}
 (\ti\beta;X_1^{\greater})
-
\!\frac{\gamma_{1k}}{\Omega_1}
\vartheta_{{\bf n}^{+}_{1k}}^{(n+1)}
(\ti\beta;P_1^{\greater})
 +\!\!\sum_{m=1}^2 n_{mk}
 \eta_{m1k}
 \frac{\gamma_{1k}}{\Omega_1}
 \vartheta_{{\bf n}^{-}_{mk}}^{(n-1)}
 (\ti\beta;P_1^{\greater})
 \Big].
\end{align}
After the second-time action it  gives rise to
\begin{align}
&\vartheta^{(n)}_{\bf n}\Big(\ti\beta;\frac{\Omega_1}{2}(
P_1^2 +  X^2_1)^{\greater}\Big)
\nl&=
\frac{1}{4\lambda_1}
\sum_{kk'}
\bigg[
\vartheta_{{\bf n}^{+\,+}_{1k1k'}}^{(n+2)}
(\ti\beta)
+
\sum_{m}
(n_{mk'}+\delta_{kk'}\delta_{m1} )\eta_{m1k'}
\vartheta_{{\bf n}^{+\,-}_{1kmk'}}^{(n)}
(\ti\beta)
+
\sum_m
n_{mk} \eta_{m1k}\vartheta_{{\bf n}^{-\,+}_{mk1k'}}^{(n)}
 (\ti\beta)
\nl & \quad
+
\sum_{mm'}
  n_{mk} \eta_{m1k}
  (n_{m'k'}-\delta_{m’m}\delta_{kk'}) \eta_{m'1k'}
  \vartheta_{{\bf n}^{-\,-}_{mkm'k'}}^{(n-2)}
 (\ti\beta)
+
\frac{\gamma_{k}\gamma_{k'}}{\Omega^2_1}
\vartheta_{{\bf n}^{+\,+}_{1k1k'}}^{(n+2)}
(\ti\beta)
\nl & \quad
-\frac{\gamma_{k}\gamma_{k'}}{\Omega^2_1}
\sum_{m}
(n_{mk'}+\delta_{kk'}\delta_{m1} )\eta_{m1k'}
\vartheta_{{\bf n}^{+\,-}_{1kmk'}}^{(n)}
(\ti\beta)
-\frac{\gamma_{k}\gamma_{k'}}{\Omega^2_1}
\sum_m
n_{mk} \eta_{m1k}\vartheta_{{\bf n}^{-\,+}_{mk1k'}}^{(n)}
 (\ti\beta)
\nl & \quad
+\frac{\gamma_{k}\gamma_{k'}}{\Omega^2_1}
\sum_{mm'}
  n_{mk} \eta_{m1k}
  (n_{m'k'}-\delta_{m'm}\delta_{kk'}) \eta_{m'1k'}
  \vartheta_{{\bf n}^{-\,-}_{mkm'k'}}^{(n-2)}
 (\ti\beta)
\bigg]
\nl
&=
\frac{1}{4\lambda_1}
\sum_{kk'}
\bigg[
\Big(
1
+
\frac{\gamma_{k}\gamma_{k'}}{\Omega^2_1}\Big)
\vartheta_{{\bf n}^{+\,+}_{1k1k'}}^{(n+2)}
(\ti\beta)
+
\Big(
1
-
\frac{\gamma_{k}\gamma_{k'}}{\Omega^2_1}\Big)
\sum_{m}
(2n_{mk}+\delta_{kk'}\delta_{m1} )\eta_{m1k}
\vartheta_{{\bf n}^{+\,-}_{1k'mk}}^{(n)}
(\ti\beta)
\nl & \quad
+
\Big(
1
+
\frac{\gamma_{k}\gamma_{k'}}{\Omega^2_1}\Big)
\sum_{mm'}
  n_{mk} \eta_{m1k}
  (n_{m'k'}-\delta_{m'm}\delta_{kk'}) \eta_{m'1k'}
  \vartheta_{{\bf n}^{-\,-}_{mkm'k'}}^{(n-2)}
 (\ti\beta)
\bigg].
\end{align}
We thus finish computing $\vartheta^{(n)}_{\bf n}(\ti\beta;{\ti H_1}^{\greater})$ in the r.h.s. of Eq.\,(23) of main text.

\clearpage
\section{More numerical details}

\subsection{Shape of control field}
\begin{figure}[h!]
\includegraphics[width=0.5\columnwidth]{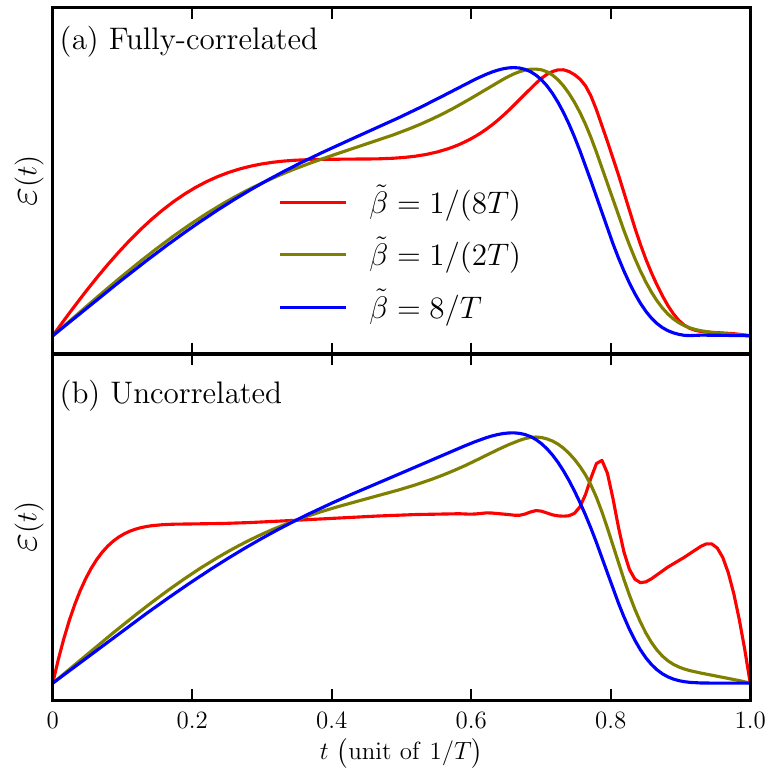}
\caption{Optimal control field $\varepsilon(t)$ for different $\ti \beta$ in the fully-correlated and uncorrelated conditions. For each curve, the field is the linear combination of the eigenvectors according to the largest three eigenvalues
of Eq.\,(15) in the main text, to make $\varepsilon(t_0)=\varepsilon(t_f)=0$. 
All the fields shown in the figure are scaled to be of the same integration strength, $I$.}\label{fig_sm1}
\label{sketch}
\end{figure}

\subsection{Behavior of variance of $F_1$}

\begin{figure}[h!]
\includegraphics[width=0.54\columnwidth]{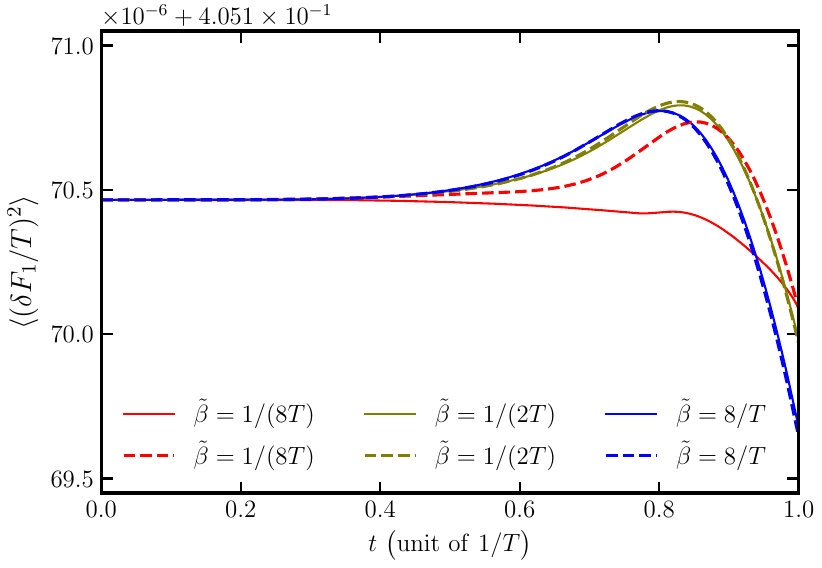}
\caption{The variance of distribution of the controlled mode $F_1$, with the solid and dashed curves (sharing the same endpoints due to the targeted control) representing the uncorrelated and fully-correlated conditions, respectively.}\label{fig_sm2}
\end{figure}

\clearpage
\subsection{Some related formulas}
We first notice the expressions reading
\bsube\label{f}
\begin{align}\label{fm}
\rho^{(0)}_{\bf 0}(t;F^{\greater}_n)=\sum_{k}
\rho^{(0)}_{\bf 0}(t; f^{\greater}_{nk} )
=
\sum_{k}
\rho^{(1)}_{{\bf 0}^{+}_{nk}}(t)
,
\end{align}
and
\begin{align}\label{fmfn}
\rho^{(0)}_{\bf 0}(t; F^{\greater}_m F^{\greater}_n)=\sum_{kk'}
\rho^{(0)}_{\bf 0}(t; f^{\greater}_{mk} f^{\greater}_{nk'})
=
\sum_{kk'}
\rho^{(1)}_{{\bf 0}^{+}_{mk}}(t; f^{\greater}_{nk'})
=
\sum_{k}
\eta_{mnk}\rho^{(0)}_{\bf 0}(t)
+
\sum_{kk'}
\rho^{(2)}_{{\bf 0}^{+\,+}_{mknk'}}
(t).
\end{align}
\esube
As a result, the variance of $F_1$ can be computed as
\begin{align} \label{ff1}
  \la (\delta F_1)^2\ra
={\rm tr}_{\tS}
\big[\rho^{(0)}_{\bf 0}(t;(F^{\greater}_1)^2)\big]- \big[{\rm tr}_{\tS}\rho^{(0)}_{\bf 0}(t; F^{\greater}_1)\big]^2
 =
\sum_{k}
\eta_{11k}
+
\sum_{kk'}
{\rm tr}_{\tS}\rho^{(2)}_{{\bf 0}^{+\,+}_{1k1k'}}
(t)
-\Big[
\sum_{k}
{\rm tr}_{\tS}\rho^{(1)}_{{\bf 0}^{+}_{1k}}(t)
\Big]^2
.
\end{align}
%
The variance of $U=F_2-F_1$ is obtained in a similar way but a bit more complicated. 
It reads
\begin{align}
\la (\delta (F_2-F_1))^2\ra=& {\rm tr}_{\tS}\big[\rho^{(0)}_{\bf 0}(t;(F^{\greater}_2-F^{\greater}_1)^2)\big]- \big[{\rm tr}_{\tS}\rho^{(0)}_{\bf 0}(t; F^{\greater}_2-F^{\greater}_1)\big]^2
\nl=&
{\rm tr}_{\tS}\big[\rho^{(0)}_{\bf 0}(t;(F^{\greater}_1)^2)\big]
+
{\rm tr}_{\tS}\big[\rho^{(0)}_{\bf 0}(t;(F^{\greater}_2)^2)\big]
-
{\rm tr}_{\tS}\big[
\rho^{(0)}_{\bf 0}(t;F^{\greater}_1 F^{\greater}_2)\big]
-
{\rm tr}_{\tS}\big[\rho^{(0)}_{\bf 0}(t;F^{\greater}_2F^{\greater}_1 )\big]
\nl&- \big[
{\rm tr}_{\tS}\rho^{(0)}_{\bf 0}(t; F^{\greater}_1)
-
{\rm tr}_{\tS}\rho^{(0)}_{\bf 0}(t;F^{\greater}_2)
\big]^2
\nl=&
\sum_{m=1}^2\sum_{n=1}^2\sum_k
(-1)^{m+n}\Big[
\eta_{mnk}
+
\sum_{k'}
{\rm tr}_{\tS}
\rho^{(2)}_{{\bf 0}^{+\,+}_{mknk'}}(t)\Big]-\Big[
\sum_{k}
{\rm tr}_{\tS}\rho^{(1)}_{{\bf 0}^{+}_{1k}}(t)
-
\sum_{k}
{\rm tr}_{\tS}\rho^{(1)}_{{\bf 0}^{+}_{2k}}(t)
\Big]^2
.
\end{align}


